\documentclass[twocolumn]{article}
\usepackage[T1]{fontenc}
\usepackage[latin1]{inputenc}
\usepackage{graphics}

\makeatletter

\newcommand{\LyX}{L\kern-.1667em\lower.25em\hbox{Y}\kern-.125emX\spacefactor1000}
\newcommand{\noun}[1]{\textsc{#1}}

\newcommand{\lyxaddress}[1]{
  \par {\raggedright #1 
  \vspace{1.4em}
  \noindent\par}
}

\makeatother

\begin{document}

\title{On a common circle: natural scenes and Gestalt rules }

\author{Mariano Sigman\protect\( ^{1,2}\protect \), Guillermo A. Cecchi\protect\( ^{1\#}\protect \),
Charles D. Gilbert\protect\( ^{2}\protect \) and Marcelo O. Magnasco\protect\( ^{1}\protect \) }

\maketitle

\lyxaddress{The Laboratories of \protect\( ^{1}\protect \)Mathematical Physics and \protect\( ^{2}\protect \)Neurobiology,
The Rockefeller University, 1230 York Avenue, New York NY 10021 \protect\( ^{\#}\protect \)Present
address: Functional Neuroimaging Laboratory, Dept. of Psychiatry, Cornell University.}

\begin{abstract}
{\footnotesize To understand how the human visual system analyzes images, it
is essential to know the structure of the visual environment. In particular,
natural images display consistent statistical properties that distinguish them
from random luminance distributions. We have studied the geometric regularities
of oriented elements (edges or line segments) present in an ensemble of visual
scenes, asking how much information the presence of a segment in a particular
location of the visual scene carries about the presence of a second segment
at different relative positions and orientations. We observed strong long-range
correlations in the distribution of oriented segments that extend over the whole
visual field. We further show that a very simple geometric rule,} \emph{\footnotesize cocircularity}{\footnotesize ,
predicts the arrangement of segments in natural scenes, and that different geometrical
arrangements show relevant differences in their scaling properties. Our results
show similarities to geometric features of previous physiological and psychophysical
studies. We discuss the implications of these findings for theories of early
vision.}{\footnotesize \par}
\end{abstract}
~

One of the most difficult problems that the visual system has to solve is to
group different elements of a scene into individual objects. Despite its computational
complexity, this process is normally effortless, spontaneous and unambiguous
\cite{1}. The phenomenology of grouping was described by the Gestalt psychologists
in a series of rules summarized in the idea of good continuation \cite{2,3}.
More quantitative psychophysical measurements have shown the existence of association
fields \cite{4} or rules that determine the interaction between neighboring
oriented elements in the visual scene \cite{5,6}. Based on these rules and
on the Gestalt ideas, pairs of oriented elements that are placed in space in
such a way that they extend on a smooth contour joining them will normally be
grouped together. 

These psychophysical ideas have been steadily gaining solid neurophysiological
support. Neurons in primary visual cortex (V1) respond when a bar is presented
at a particular location and at a specific orientation \cite{7}. In addition,
the responses of V1 neurons are \emph{modulated} by contextual interactions
\cite{6,8,9,10,11,12,13,14,15}, such as the joint presence of contour elements
within the receptive field and in its surround. This modulation depends upon
the precise geometrical arrangement of linear elements \cite{6,16}, in a manner
corresponding to the specificity of linkage of cortical columns by long-range
horizontal connections \cite{17,18}. Thus neurons in V1 interact with one another
in geometrically meaningful ways, and through these interactions, neuronal responses
become selective for combinations of stimulus features that can extend far from
the receptive field core.

The rules of good continuation, the association field, and the connections in
primary visual cortex provide evidence of interaction of pairs of oriented elements
at the psychophysical, physiological and anatomical level. The nature of the
interaction is determined by the geometry of the arrangement, including spatial
arrangement and the orientation of segments within the visual scene. An important
question is whether this geometry is related to natural geometric regularities
present in the environment. It is well known that natural images differ from
random luminance distributions \cite{19,20} but the structural studies of natural
scenes have not yet addressed the existence of geometrical regularities. Here,
we address this issue, by studying whether particular pairs of oriented elements
are likely to co-occur in natural scenes as a function of their orientation
and relative location in space. Our results are focused on two different aspects
of the organization of oriented elements in natural scenes: scaling and geometric
relationships. We will show that these two are interdependent.

Scaling measurements involve studying how the probability of finding a co-occurring
pair changes as a function of the relative distance. A classic result in the
analysis of natural scenes is that the luminance of pair of pixels is correlated
and that this correlation is scale invariant \cite{19,20}. This indicates that
statistical dependencies between pair of pixels do not depend on whether the
observer zooms in on a small window or zooms out to a broad vista. The scale
invariance results from stable physical properties such as a common source of
illumination and the existence of objects of different sizes and similar reflectance
properties \cite{21}. Here we show that for particular geometries, the probability
of finding a pair of segments follows a power law relation and thus is scale
invariant. We show further that a very simple geometric rule, consistent with
the idea of good continuation, predicts the arrangement of segments in natural
scenes.

\subsubsection*{Materials and Methods }

Images were obtained from a publicly available database (\emph{http://hlab.phys.rug.nl/imlib/index.html})
\cite{22} of about 4000 uncompressed black and white pictures, 1536x1024 pixels
in size and 12 bits in depth, with an angular resolution of \~{}1 minute of
arc per pixel. This particular database was chosen because of the high quality
of its pictures, especially the lack of motion and compression artifacts, which
would otherwise overwhelm our statistics. To obtain a measure of local orientation
we used the steerable filters of the \( H_{2} \) and \( G_{2} \) basis \cite{23}.
Using steerable filters, the energy value at any orientation can be calculated
by extrapolating the responses of a set of basis filters. A \( G_{2} \) filter
is a second derivative of a Gaussian and the \( H_{2} \) filter is its Hilbert
transform. \( H_{2} \) and \( G_{2} \) filters have the same amplitude spectra,
but they are \( 90^{\circ } \) out of phase; that makes them a quadrature pair
basis filters. The size of the filters used was 7x7 pixels. A measure of oriented
energy was obtained by combining both sets of filters \( E(\varphi )=G^{2}_{2}(\varphi )+H^{2}_{2}(\varphi ) \)
\cite{23}. This measure is repeated at every pixel of the image to obtain the
energy function for each image (\emph{n}) of the ensemble \( \{E_{n}(x,y,\varphi )\} \).
To study the joint statistics of \( E(x,y,\varphi ) \), we discretized the
different orientations at 16 different values, \( 0=(\frac{-\pi }{32},\frac{\pi }{32}),\, 1=(\frac{\pi }{32},\frac{3\pi }{32}),\, ...,\, 15=(\frac{29\pi }{32},\frac{31\pi }{32}) \)
as shown in the color representation of orientations of Fig. 1. With this information
one can obtain a measure of the statistics of pairs of segments by calculating
the correlation (weighting the co-occurrences of segments by their energy)
\[
C(\Delta x,\Delta y,\varphi ,\psi )=\qquad \qquad \qquad \qquad \qquad \]

\[
\frac{1}{N}\sum ^{N}_{n=1}\int \int E_{n}(x,y,\varphi )E_{n}(x+\Delta x,y+\Delta y,\psi )dxdy\]
where \emph{N} is the total number of images and the integral is over each of
the images of the ensemble. We were interested in measuring long-range correlations
so we studied values of \( \Delta x,\Delta y=\{-256,256\} \). The correlation
matrix has dimensions 512x512x16x16 and each point results from averaging 4000
integrals over a 1536x1024 domain. To simplify the computations, for the general
case, we decided to store at each pixel, for every image, the maximum energy
value \( E(\varphi _{max}) \) and its corresponding orientation \( \varphi _{max} \).
An energy threshold \( E_{T} \) was arbitrarily set to match the visual perception
of edges in a few images. Pixels in an image were considered ``oriented''
if \( E(\varphi _{max})\geq E_{T} \), ``non-oriented'' otherwise. This unique
threshold value was applied to all images in the ensemble. Thus, for each image,
we extracted a binary field \( E_{n}^{bin}(x,y)=\{0,1\} \) and an orientation
field \( Ang_{n}(x,y)=\{1,...,16\} \). From this binary field we can construct
a histogram of co-occurrences: how many times an element at postion \emph{(x,y)}
was considered oriented with orientation \( \varphi  \) and at position \( (x+\Delta x,y+\Delta y) \)
a segment was considered oriented with orientation \( \psi  \). Thus, formally,
the histogram is obtained as \( C \), taking as the Energy function \( E_{n}(x,y,\varphi )=1 \)
if \( \varphi =Ang_{n}(x,y) \) and \( E_{n}^{bin}(x,y)=1 \); \( E_{n}(x,y)=0 \)
in any other case. The computation is reduced to counting the co-occurrences
in the histogram \( H(\Delta x,\Delta y,\varphi ,\psi ) \) with \( \Delta x=\{=256,256\},\! \Delta y=\{-256,256\},\! \varphi ,\psi =(0,\pi /16,2\pi /16,...,\pi ) \).
From the histogram we obtained a measure of statistical dependence. While choosing
the threshold followed computational reasons, cortical neurons perform a thresholding
operation and thus the measure of linear correlation (weighting co-occurrences
by their energy) is not necessarily a more accurate measure of statistical dependence.
The histogram was used for all the data shown in Fig. 2a, 2b and 2c, Fig. 3,
Fig. 4, and Fig. 5. For Fig. 2d, for the particular case of collinear interactions
we computed the full linear cross-correlation. This computation is considerably
easier since it is done for fixed values of orientation and direction in space.
The two measures shown (Laplacian correlation and collinear correlation) were
obtained according to the formulas: 
\[
C(r)=\qquad \qquad \qquad \qquad \]
 
\[
\sum _{x,y}\sum _{\Delta x^{2}+\Delta y^{2}=r^{2}}E^{Lap}(x,y)E^{Lap}(x+\Delta x,y+\Delta y)\]

\[
\qquad \qquad \qquad \qquad \qquad -(\sum _{x,y}E^{Lap}(x,y))^{2}\]
 for Laplacian filtering, and
\[
C(r)=\sum _{x,y}E(x,y,0)E(x+r,y,0)-(\sum _{x,y}E(x,y,0))^{2}\]
for collinear oriented filtering. 

A quantitative signature of scale invariance is given by a function of the form
\( C=r^{-a} \) (power law) where \emph{C} is the correlation, \emph{r} the
distance and \emph{a} constant. If the scale is changed \( r\rightarrow \lambda r=r' \)
the function changes as \( C(r)=\lambda ^{-a}r^{-a}=kC(r') \) where \emph{k}
is a constant. A power law is easily identified as a linear plot in the log-log
graph, which is clear from the relation \( \log (C)=-a\log (r) \). 

The axis of maximal correlation (Fig. 5b) was calculated as follows. For each
pair of orientations \( (\varphi ,\psi ) \), a measure of co-occurrence was
calculated integrating across 16 different lines of angles of values \( (0,\frac{\pi }{128},\frac{2\pi }{128},...,\pi ) \)
over distances of {[}-40,40{]} of the center of the histogram. Thus, for an
angle \( \theta  \) and orientations \( (\varphi ,\psi ) \) the measure of
co-occurrence is: \( P_{\varphi ,\psi }(\theta )=\sum ^{40}_{i=-40}H(\cos (\theta )*i,\sin (\theta )*i,\varphi ,\psi ) \).
We then calculated the direction of maximal correlation \( \theta _{max}(\varphi ,\psi ) \)
and we grouped all angles with common relative orientation, \( \varphi -\psi =\xi  \).
We had 16 different values for each \( \xi  \) and from these 16 different
values we calculated the mean \( P(\theta ,\epsilon )=<\theta _{max}(\psi ,\psi +\epsilon )>_{\psi } \)
and the standard error. To calculate the mean energy as a function of relative
orientation (Fig. 3) we integrated the histogram in spatial coordinates for
each pair of orientations in space, and, as before, the different pairs where
grouped according to their relative difference in orientation to calculate a
mean value and a standard deviation, \( E_{\varphi ,\psi }=\int ^{100}_{x=-100}\int ^{100}_{y=-100}H(x,y,\varphi ,\psi )dxdy \)
and \( E(\varphi )=\langle E_{\epsilon ,\epsilon +\varphi }\rangle _{\epsilon } \).
The code was parallelized using MPI libraries and run over a small Beowulf cluster
of Linux workstations.

In general, horizontal and vertical directions had better statistics since there
are more horizontal or vertical segments than oblique in the images; these special
orientations are also the ones most prone to artifacts from aliasing, staircasing,
and the ensemble choice. Since we are interested in this study in the correlations
as a function of relative distance and orientations, all the quantitative measurments
were performed averaging over all orientations. However, the results shown still
held true for each individual orientation.

\subsubsection*{Results }

All 4000 images used in this study were black and white, 1536x1024 pixels in
size and 12 bits in depth. We used a set of filters to obtain a measure of orientation
at each pixel of every image of the database \cite{23}. The filters were 7x7
pixels in size and thus provided a local measure of orientation. The output
of the filter was high at pixels where contrast changed abruptly in a particular
direction, typically by the presence of line segments or edges, but also corners,
junctions or other singularities (Fig. 1). If the output of the filters were
statistically independent then we would expect a flat correlation as a function
of \( (\Delta x,\Delta y,\varphi ,\psi ) \). In polar coordinates \( (r,\theta ,\varphi ,\psi ) \),
the two problems which we address are naturally separated: the scaling properties
result from studying how the histogram depends on \emph{r} (distance) whereas
the geometry does it from the dependence of the histogram on \( \theta  \),
\( \varphi  \) and \( \psi  \). 

We studied the number of co-occurring pairs of segments as a function of their
relative distance for different geometries (Fig 2a,b,c). The different geometric
configurations correspond to the different orientations of the segments and
their relative position within an image. We first studied the number of co-occurrences
as a function of distance in the line spanned by the orientation of the reference
segment, averaged across all possible orientations of the reference line (Fig
2a). When both segments have the same orientation we observe a scale invariant
behavior, indicated by a linear relationship in the log-log plot (see methods).
Also it can be seen from this plot that collinear co-occurrences are more frequent
than any other configuration. Fig. 2b shows the probability of co-occurrences
is higher for the vertical orientation, and scale invariance extends over a
broader range. 

The scaling properties are qualitatively different for segments positioned side-by-side,
along a line orthogonal to the orientation of the first segment (Fig. 2c). Iso-oriented
pairs were again the most frequent, but their co-occurrence in the orthogonal
direction to the orientation of the first segment (fig 2c, black line) does
not appear to be scale invariant. This is reflected by the presence of a kink
as opposed to a straight line (power law) in the log-log plot, indicating well-defined
scales with different behaviour. 

It is worth comparing the scale of interactions one observes by using different
kinds of filters. Before filtering images, the luminance shows correlations,
which follows a power law behavior \cite{19,20}. After applying a Laplacian
filter (equivalent to a center-surround operator, which measures non-oriented
local contrast), the image is mostly decorrelated (Fig. 2d, red circles) \cite{24,25}.
This is seen in the exponential decay of the correlations, and the fact that
the correlations show similar behavior after a pixel-by-pixel shuffling of the
image (Fig. 2d, cyan circles). The strength and scaling of the correlations
across the collinear line changes radically when one uses an oriented filter.
In this example, to make a direct comparison between the various filters, we
weighted each pair of segments by their energy value (linear cross-correlation,
instead of applying a threshold as was done in the earlier calculations). This
calculation was done for the vertical reference line orientation, which showed
long-range correlations (Fig. 2b, black circles), over much longer distances
than observed with the Laplacian filter. Moreover, these correlations were not
present when measured in the shuffled images (Fig. 2, green circles). It is
clear from the above analysis that when one uses oriented filters one reveals
strong correlations extending over large distances. The next question is how
these correlations depend on the relative orientation of the line elements,
and whether these dependencies have any underlying geometry. We first calculated
the total number of co-occurrences as a function of the relative difference
in orientation. Co-occurrences decreased as the relative orientation between
the pair of segments increased, being maximal when they were iso-oriented and
minimal when they were perpendicular (Fig. 3). 

The next observation concerns spatial structure. The probability of finding
co-occurring pair of segments was not uniform but rather displayed a consistent
geometric structure. If the two segments were iso-oriented, their most probable
spatial arrangement was as part of a common line, the collinear configuration
(Fig. 4a). As the relative difference in orientation between the two segments
increased, two effects were observed. The main lobe of the histogram (which
in the iso-oriented case extends in the collinear direction) rotated and shortened,
and a second lobe (where co-occurrences were also maximized) appeared at \( 90^{\circ } \)
from the first (Fig 4a-e). This effect progressed smoothly until the relative
orientation of the two segments was \( 90^{\circ } \), where the two lobes
were arranged in a symmetrical configuration, lying at \( 45^{\circ } \) relative
to the reference orientation. Thus, pairs of oriented segments have significant
statistical correlations in natural scenes, and both the average probability
and spatial layout depend strongly upon their relative orientation. Remarkably,
the structure of the correlations followed a very simple geometric rule. A natural
extension of collinearity to the plane is cocircularity. While two segments
of different orientations cannot belong to the same straight line, they may
still be tangent to the same circle if they are tilted at identical but opposite
angles to the line joining them. Given a pair of segments tilted at angles \( \psi  \)
and \( \varphi  \) respectively, they should lie along two possible lines,
at angles \( (\varphi +\psi )/2 \) or \( (\varphi +\psi +\pi )/2 \), in order
to be cocircular (Fig. 4f). This is the arrangement we observed in natural scenes.
The measured correlations, given any relative orientation of edges, were maximal
when arranged along a common circle. To quantify this we calculated the orientation
of the axis where co-occurrences were maximal. We did that for different relative
orientations and compared it to the value predicted by the cocircularity rule
(Fig. 5). This is particularly remarkable because the comparison is not a fit,
since the cocircularity rule has no free parameters.

\subsubsection*{Discussion }

We have shown that there are strong, long-range correlations between local oriented
segments in natural scenes, that their scaling properties change for different
geometries, and that their arrangement obeys the cocircularity rule. The filters
we used for edge detection in our images were an oriented version of Laplacian-like
filters, in that they were local but had elongated, rather than circularly symmetric,
center-surround structures. This change is analogous to the difference between
filters in the LGN and simple cells in the primary visual cortex. Thus, given
that Laplacian filtering decorrelates natural scenes \cite{24}, it was surprising
to find the long-range correlations and scale invariant behavior of the collinear
configuration. It is important to remark that our measure of correlation does
not differ only on the type of filters used (elongated vs. circular symmetric)
but also on the fact that we measured the correlations along a line containing
the pair of segments. Long contours are part of the output of the Laplacian
filters and thus the image should show correlations which might be hidden when
integrating them across an area, essentially because a curve has zero area and
thus the correlations along a curve are not significant when integrated over
the two-dimensional field of view. The findings of long-range correlations of
oriented elements extends the notion that the output of linear local oriented
filtering of natural scenes cannot be statistically independent \cite{26} and
shows that those correlations might be very significant through global portions
of the visual field for particular geometries. 

The cocircular rule has been used heuristically to establish a pattern of interactions
between filters in computer vision \cite{1,27,28,29}, and psychophysical studies
suggest that the human visual system utilizes a local grouping process, ``association
field'', with a similar geometric pattern \cite{4}. Our finding provides an
underlying statistical principle for the establishment of form and for the Gestalt
idea of good continuation, which states that there are preferred linkages endowing
some contours with the property of perceptual saliency \cite{2}. An important
portion of the classical Euclidean geometry has been constructed using the two
simplest planar curves, the line and the circle \cite{30}; here we show that
those are, in the same order, the most significant structures in natural scenes.

We have reported the emergence of robust geometric and scaling properties of
natural scenes. This raises the question of the underlying physical processes
that generate these regularities. While our work was solely based on statistical
analysis, we can speculate on the possible constraints imposed by the physical
world.  In a simplyfing view, we can think of a natural image as composed by
object boundaries or contours, and  textures.  Collineal pairs of segments are
likely to belong to a common contour;  thus, our finding of scale invariance
for collineal correlations is in agreement with the idea that scale-invariance
in natural images is a consequence of the distribution of  apparent sizes of
objects \cite{21}. Parallel segments, on the contrary, may be part of a common
contour as well as a common texture, which would explain the two scaling regimes
we observed.  Cocircularity in natural scenes probably arises due to the continuity
and smoothness of object boundaries; when averaged over objects of vastly different
sizes present in any natural scene, the most probable arrangement for two edge
segments is to lie on the smoothest curve joining them, a circular arc. These
ideas, however, requiere an investigation which is beyond the scope of this
paper.  

The geometry of the pattern of interactions in primary visual cortex parallels
the interactions of oriented segments in natural scenes. Long-range interactions
tend to connect iso-oriented segments \cite{17,18} and interactions between
orthogonal segments, which span a short range in natural scenes, may be mediated
by short-range connections spanning singularities in the orientation and topographic
maps in the primary visual cortex \cite{31}. The finding of a correspondence
between the interaction characteristics of neurons in visual cortex and the
regularities of natural scenes suggests a possible role for cortical plasticity
early in life, in order for the cortex to assimilate and represent these regularities.
This plasticity might be mediated by Hebbian-like processes, reinforcing connections
on neurons whose activity coincide, i.e., their corresponding stimuli are correlated
under natural visual stimulation. Such plasticity could extend to adulthood
to accommodate perceptual learning of novel and particular forms \cite{32}.

While we find coincidences between the pattern of interactions in V1 and the
distribution of segments in natural scenes, the sign of the interactions plays
a crucial role. Reinforcing or facilitation of co-occurring stimuli (positive
interaction) results in Hebbian-like coincidence detectors, while inhibiting
the response results in Barlow-like detectors of ``suspicious coincidences''
which ignore frequent co-occurrences \cite{33}. Interestingly the Hebbian idea
and the decorrelation hypothesis represent two sides of the same coin. From
our measurements of the regularities in natural scenes, and previous studies
on the higher order receptive field properties in primary visual cortex, it
appears that both type of operations exist. The response of a cell in V1 is
typically inhibited when a second flanking segment is placed outside of its
receptive field along an axis orthogonal to the receptive field orientation.
This interaction is referred as side-inhibition, which is strongest when the
flanking segment has the same orientation as the segment inside the receptive
field \cite{13,15,34}. In the present study we found that iso-orientation is
the most probable arrangement for side-by-side segments in natural scenes, which
therefore constitutes an example, in the domain of orientation, of decorrelation
through inhibition. This inhibition may mediate the process of texture discrimination
\cite{13,16,35}. The property of end-inhibition has also been interpreted as
a mechanism to remove redundancies and achieve statistical independence \cite{36}.
The finding that responses of V1 neurons are sparse when presented with natural
stimuli \cite{37} and models of normalization of neuronal responses in V1 tuned
to the statistics of natural scenes \cite{26} also support the idea that the
interactions in V1 play an important role in decorrelating the output from V1.
This is consistent with the general idea that one of the important functions
of early visual processing is to remove redundant information \cite{38,39,40}
and suggests that interactions in V1 may continue with the process of decorrelation
which is achieved by Laplacian \cite{24} and local oriented filtering \cite{41,42}. 

But the visual cortex also can act in the opposite way, reinforcing the response
to the most probable configurations. This is seen in the collinear configuration,
which is the one that elicits most facilitation, and therefore illustrates how
V1 can enhance the regularities in natural scenes. The fact that those correlations
are significant over the entire visual field and are highly structured suggests
that this is not a residual, or second-order process. The opposing processes
of enhancement of correlations and decorrelation may be mediated by different
receptive field properties, which can exist within the same cell. The same flank
can inhibit or facilitate depending on the contrast \cite{43,44} suggesting
that V1 may be solving different computational problems at different contrast
ranges or different noise-to-signal relationship. The dialectic behavior of
visual cortex shows that the interplay between decorrelation (extraction of
suspicious coincidences) and enhancement of particular set of regularities (identification
of form) may be mediated by the same population of neurons. While the decorrelating
process may be required to operate in the orientation domain to solve the problem
of texture segmentation, particular sets of coincidences, which are repeated
in the statistics, such as the conjunction of segments that form contours, need
to be enhanced in the process of identification of form.

\subsection*{Acknowledgements}

We thank M. Kapadia for suggesting connections of our work with neurophysiological
data, and D.R. Chialvo, R. Crist, A.J.Hudspeth and A. Libchaber for constructive
comments on the manuscript. We thank specially P. Penev for stimulating input
in early stages of the project. Supported by NIH grant EY 07968 and by the Winston
(GC) and Mathers Foundations (MM) and the Burroughs Wellcome Fund (MS).

\section*{Figure Captions}

~

\vspace{0.3cm}
{\centering \fbox{\rule[-0.5in]{0pt}{1in}empty figure path} \par}
\vspace{0.3cm}

\noun{Figure 1}. \textbf{}An example of the filtering process we applied to
an image. a) The original image. b) The image after processing with local oriented
filters \cite{23}. The maximal orientation was calculated at each point. The
image was converted to binary by considering \char`\"{}oriented\char`\"{} only
the pixels that after being filtered at their maximal orientation exceeded a
given threshold. In the figure, the maximal orientation is shown using a color
code. 

\vspace{0.3cm}
{\centering \resizebox*{0.5\textwidth}{!}{\rotatebox{-90}{\includegraphics{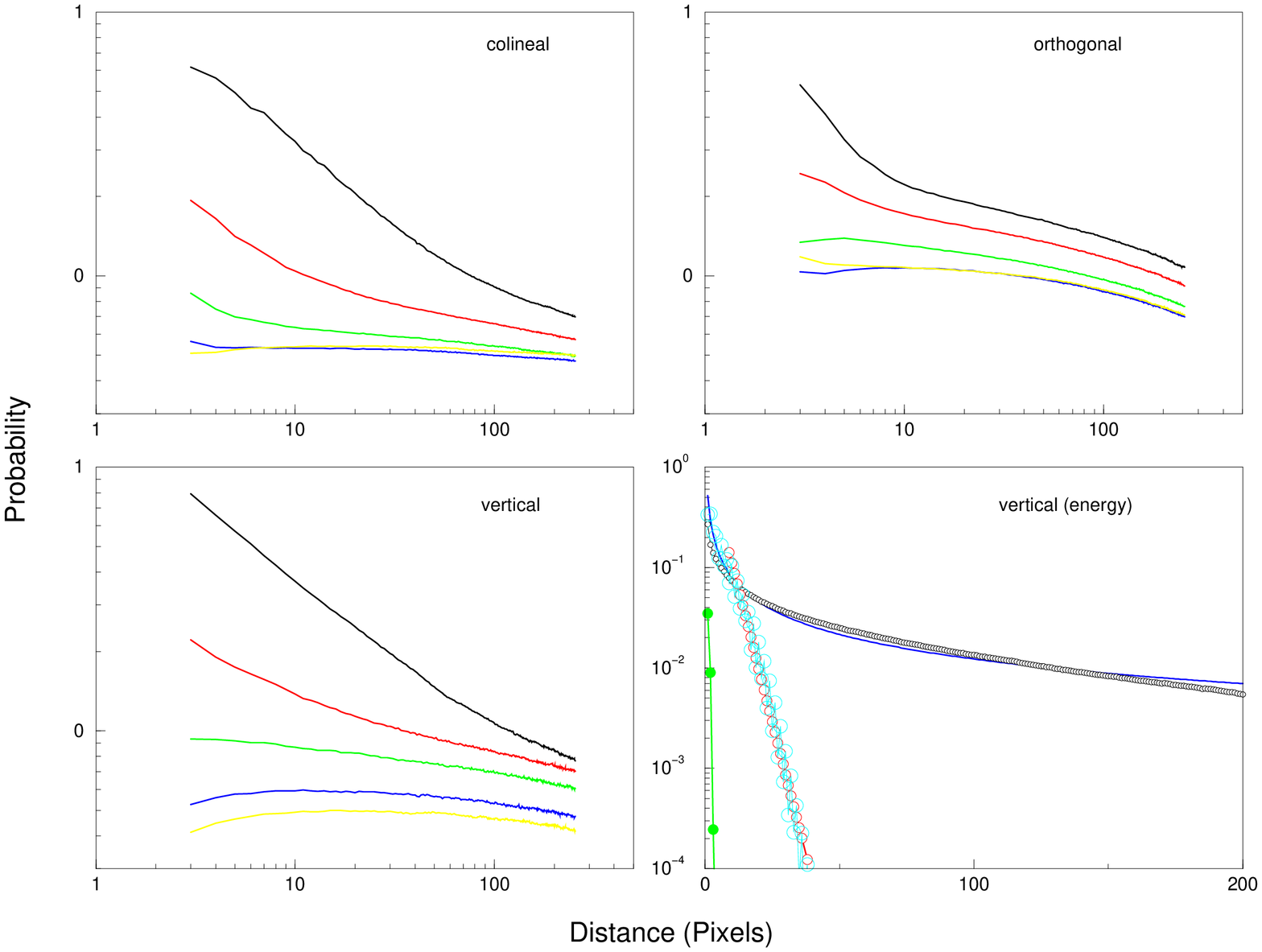}}} \par}
\vspace{0.3cm}

\noun{Figure 2.} Scaling behaviors for different geometrical configurations.
a) The number of co-occurrences between two segments in the relative positions
within the line that the orientation of the first segment spans is shown for
different orientations of the second segment. This measure was averaged over
all possible orientations of the first segment. The collinear configuration
is the most typical case and displays a scale invariant behavior as indicated
by the linear relationship in the log-log plot. b) The strength of the correlations
and the degree to which it can be approximated to a power law are more pronounced
for the particular case in which the reference line segment is vertical. c)
The same measure when the two segments are at a line \( 90^{\circ } \) apart
from the orientation of the first segment. In all three cases, black corresponds
to iso-orientation, red to \( 22.5^{\circ } \) respect to the first segment,
green to \( 45^{\circ } \), blue to \( 67.5^{\circ } \) and yellow to \( 90^{\circ } \).
d) Full cross-correlation as a function of distance for Laplacian filtering
(red circles), oriented filters in the collinear vertical direction (black circles)
and for both cases after shuffling the images. The Laplacian filtered image
is de-correlated, as can be seen from the fact that it shows the same structure
as its shuffled version (cyan circles). Collinear configuration shows long-range
correlations, which follow a power law of exponent 0.6 (blue line, y=x\( ^{-0.6} \))
and are not present when the image is shuffled (green circles).

\vspace{0.3cm}
{\centering \resizebox*{0.5\textwidth}{!}{\rotatebox{-90}{\includegraphics{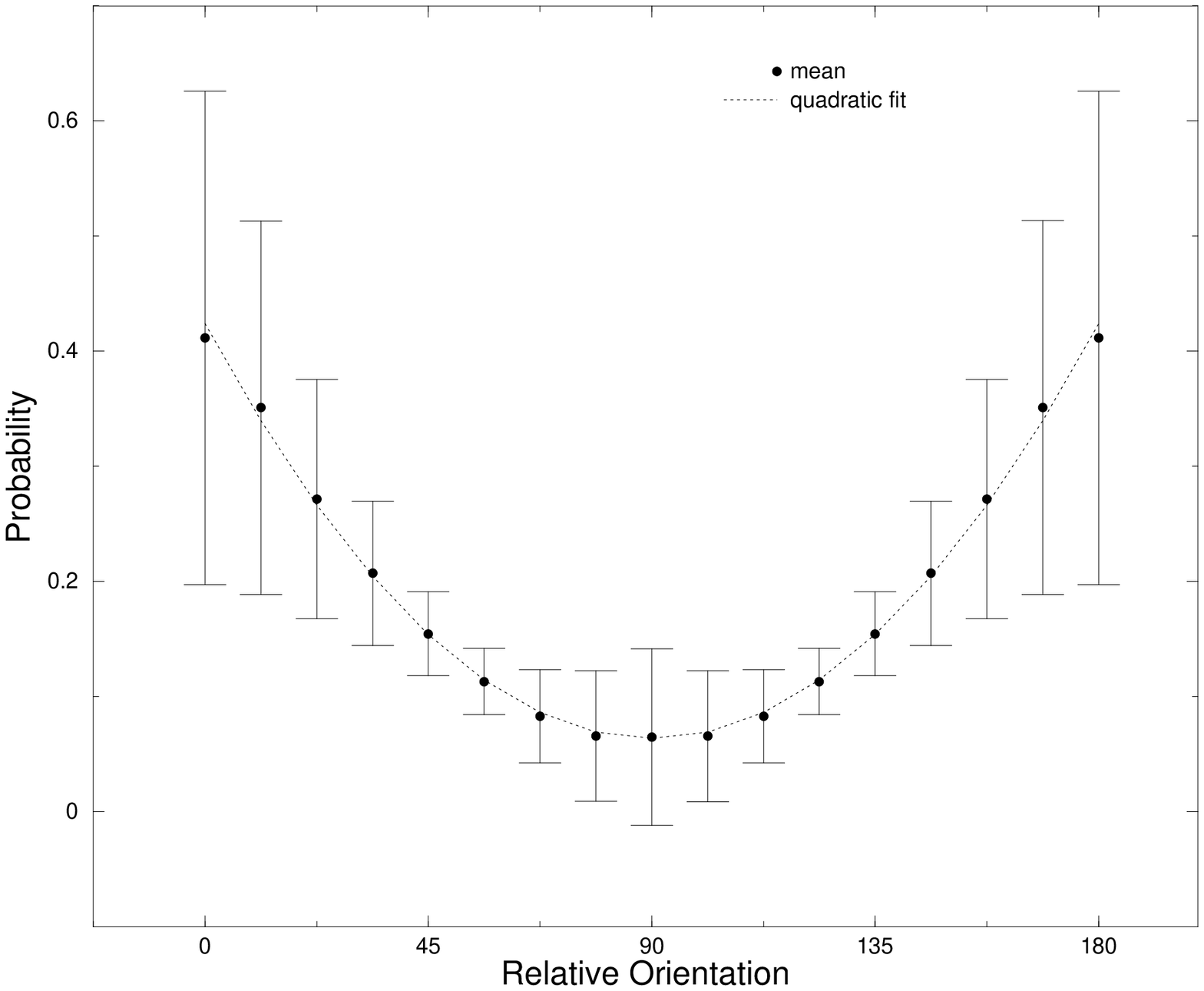}}} \par}
\vspace{0.3cm}

\noun{Figure 3}. The number of co-occurring pairs of segments as a function
of their relative difference in orientation \( (\xi =\psi -\varphi ) \). These
values where obtained after integrating the histogram of co-occurrences in space
for different angular configurations. Each point in the graph \( (\xi ) \)
corresponds to the average and the standard deviation of the 16 different configurations
obtained by choosing one of the 16 possible values for the first orientation
\( (\varphi ) \) and then setting \( \psi =(\varphi +\xi )(modulo\! \! \! 16) \). 

\vspace{0.3cm}
{\centering \fbox{\rule[-0.5in]{0pt}{1in}empty figure path} \par}
\vspace{0.3cm}

\noun{Figure 4}. Plot of the spatial dependence of the histogram of co-occurring
pairs for different geometrical configurations. a) The probability of finding
a pair of iso-oriented segments as a function of their relative position (b),
a pair of segments at relative orientation of \( 22.5^{\circ } \), (c) \( 45^{\circ } \),
(d) \( 67.5^{\circ } \) or (e) \( 90^{\circ } \). f) Cocircularity solution
for a particular example of two segments. The solutions to the problem of cocircularity
are two orthogonal lines, whose main have values \( (\psi +\varphi )/2 \) or
\( (\psi +\varphi +\pi )/2 \). For the example given, \( \varphi  \)(red segment)\( =20^{\circ } \),
\( \psi  \)(blue segment)\( =40^{\circ } \) and the two solutions (green lines)
are \( 30^{\circ } \) and\( 120^{\circ } \) (all angles from the vertical
axis). 

\vspace{0.3cm}
{\centering \resizebox*{0.5\textwidth}{!}{\rotatebox{-90}{\includegraphics{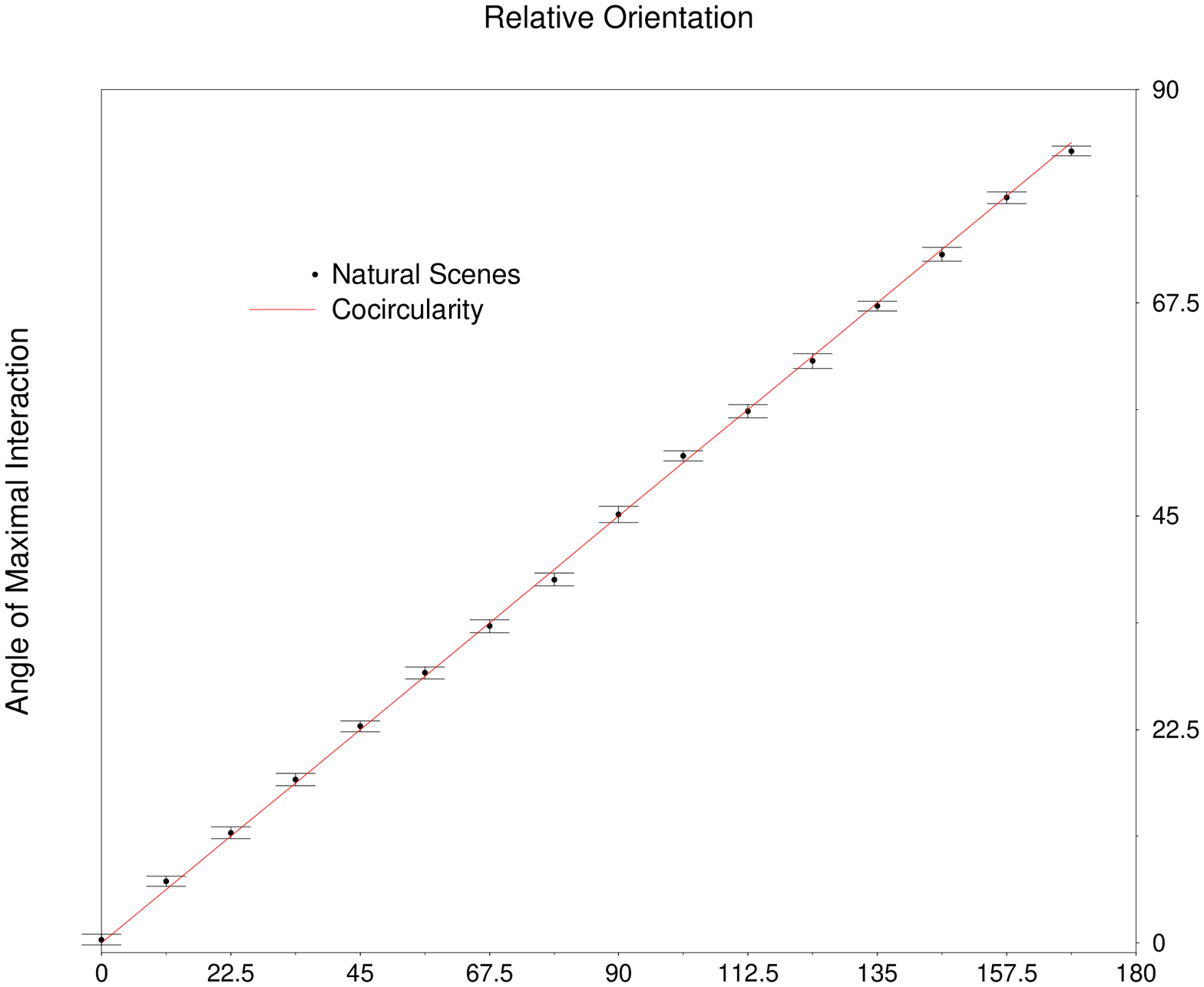}}} \par}
\vspace{0.3cm}

\noun{Figure 5}. Quantitative analysis of the spatial maps. Orientation of
the axis where co-occurring pairs of oriented elements of relative orientation
\( (\xi =\psi -\varphi ) \) are maximized. The axis of maximal probability
was calculated relative to the orientation of the segment in the center \( (\psi ) \).
This was done for the 16 possible orientations of ( and the corresponding values
of \( \varphi =(\psi +\xi )(modulo\! \! 16) \), and we computed for each (
the mean and standard error. The solid line corresponds to the solution predicted
by the cocircular rule.

~

\end{document}